\documentclass[USletter,10pt,twoside]{article}

\usepackage[utf8]{inputenc}
\usepackage[english]{babel}
\usepackage[T1]{fontenc}
\usepackage{url}
\usepackage[colorlinks=true, allcolors=blue]{hyperref}
\usepackage{graphicx}
\usepackage[left=85pt, right=85pt, top=80pt, bottom=95pt,
headheight=15pt,%
headsep=10pt]{geometry}%
\usepackage{amsmath, amssymb}
\usepackage{authblk}
\usepackage{algorithm, algpseudocode}
\usepackage{mathptmx}

\usepackage[figurename=Fig.]{caption}

\title{Inverse design of gradient-index volume multimode converters}

\author[1,*]{Nicolas Barré}
\author[1]{Alexander Jesacher}

\affil[1]{Institute of Biomedical Physics, Medical University of
  Innsbruck, M\"{u}llerstra\ss e 44, 6020 Innsbruck, AT}

\affil[*]{Corresponding author: nicolas.barre@i-med.ac.at}

\begin{document}

\maketitle

\begin{abstract}
  Graded-index optical elements are capable of shaping light precisely
  and in very specific ways. While classical freeform optics uses only
  a two-dimensional domain such as the surface of a lens, recent
  technological advances in laser manufacturing offer promising
  prospects for the realization of arbitrary three-dimensional
  graded-index volumes, i.e. transparent dielectric substrates with
  voxel-wise modified refractive index distributions.  Such elements
  would be able to perform complex light transformations on compact
  scales.

  Here we present an algorithmic approach for computing 3D
  graded-index devices, which utilizes numerical beam propagation and
  error reduction based on gradient descent.  We present solutions for
  millimeter-sized elements addressing important tasks in photonics: a
  mode sorter, a photonic lantern and a multimode intensity beam
  shaper. We further discuss suitable cost functions for all designs
  to be used in the algorithm.  The 3D graded-index designs are
  spatially smooth and require a relatively small refractive index
  range in the order of $10^{-2}$, which is within the reach of direct
  laser writing manufacturing processes such as two-photon
  polymerization.
\end{abstract}

\section{Introduction}

Light has become one of the most powerful and versatile tools. It is
an important medium for transporting information, either between
communicating partners or from an object under investigation to a
detector in imaging applications. Light also plays a key role in
laser-based manufacturing processes, such as welding, cutting and
direct laser lithography.

However, as recently highlighted~\cite{rubinsztein2016roadmap,
  forbes2021structured, piccardo2021roadmap}, exploiting the full
physical potential of light requires solutions for its highly specific
and custom shaping.  For example, increasing the telecommunication
bandwidth demands directing photonic signals from many single mode
fibers into a single multi-core or multimode fiber with high
efficiency~\cite{puttnam2021space}.  Likewise, efficiently exploiting
the high optical power of industrial lasers requires to reshape their
often unsuitable native mode
profiles~\cite{li2015high,ackermann2021uniform}.

In any case, the restructuring of light should ideally take up only
little space.  To this end, the processing of dielectric materials
such as glasses or polymers with ultra-fast lasers has opened
promising routes towards the creation of three-dimensional,
miniaturized light shapers~\cite{gross2015ultrafast}.  A variety of
devices such as photonic lanterns, beam combiners and mode
multiplexers made from glass-embedded waveguide arrangements have been
successfully demonstrated and are making their way towards
commercialization. Such waveguides are formed by translating a
femtosecond laser focus through a glass volume along the desired
guiding paths, along which the refractive index of the glass is
permanently modified.  Whilst undoubtedly powerful, devices using
waveguides as ``building blocks'' are nevertheless limited in the
sense that they don't fully exploit all degrees of freedom of a
volume.  Ideally, one would like to have the power of arbitrarily
changing the entire three-dimensional refractive index distribution at
the microscale.  Progress towards the fabrication of such 3D gradient
index materials for optical wavelengths has been recently reported
using two-photon polymerization~\cite{vzukauskas2015tuning,
  ocier2020direct} and 3D printed glass optics~\cite{dylla20203d}.
Likewise, the computational design of 3D gradient index optics in
order to shape arbitrary transverse beam profiles is continuously
advancing.  In Ref.~\cite{kunkel2020numerical}, the authors use an
irradiance mapping scheme relying on geometric optics, which is found
by employing algorithms used in optimal mass transport problems, on
top of which they run a multiplane Gerchberg-Saxton-like algorithm in
order to account for diffraction.  The proposed algorithm, however,
has some constraints with regards to the smoothness and continuity of
input/output functions. Furthermore, it has so far only been
demonstrated for single-mode shaping, that is sculpting a specific
coherent output field from a single coherent input beam.

{Here we propose a new algorithmic approach for the design of 3D
  gradient index devices. Unlike previous methods for designing
  gradient index or freeform optics, we refrain from using tools used
  in ray-optical designs.  Instead, we employ an approach related to
  machine learning, i.e., numerical beam propagation in conjunction
  with suitable cost functions that are optimized using error
  back-propagation and gradient descent.  We demonstrate the
  effectiveness and versatility of our approach by computing a variety
  of highly miniaturized 3D gradient index designs that perform key
  tasks in integrated photonics.  After a detailed explanation of our
  computational method and suitable cost functions in
  section~\ref{algorithm}, we present solutions for three application
  cases in section~\ref{results}.}

Firstly, we find a solution for a mode sorter of only 2.5~mm length,
which is capable of specifically mapping 45 Gaussian input beams onto
the Hermite-Gaussian modes of a multimode fiber with an average
conversion efficiency of 98.6\%.  Secondly, we design an equally small
photonic lantern, which couples light from 45 Gaussian inputs
unspecifically into a multimode fiber with 99.2\% efficiency.
Finally, we present a highly miniaturized beam shaper of only 0.5~mm
length, which turns the irradiance profile of a monochromatic
$50\,\textrm{\textmu m}$-diameter multimode fiber source into a
speckle-free square of $60\,\textrm{\textmu m}$ side length.

All presented designs exhibit smooth 3D refractive index modulations
within a dynamic range of about $10^{-2}$ and are therefore feasible
for experimental realization using 2-photon
polymerization~\cite{vzukauskas2015tuning}.

\section{Inverse design algorithm}\label{algorithm}

In the following, we present an algorithm that allows to design
gradient-index structures for the conversion of a set
$\{u_n\}_{1\leq n \leq N}$ of $N$ mutually incoherent transverse input
modes to another set of output modes $\{v_n\}_{1\leq n \leq N}$,
according to various cost functions depending on the application.

We follow a typical inverse design approach by first defining a
forward model propagating the input modes from their definition plane
$z_{in}$ to the destination plane $z_{out}$, through a structure of
refractive index contrast $\Delta\textrm{RI}(x,y,z)$ that is initially
unknown and will be updated iteratively. We then define different cost
functions for several classes of multimode beam shaping problems and
present a generic error backpropagation algorithm to compute
$\Delta\textrm{RI}$ gradients. Finally, we combine these different
stages into a simple gradient descent iteration scheme to solve
$\Delta\textrm{RI}$ and eventually apply some constraints to it.

\subsection{Forward model}

\begin{figure}[ht]
    \centering
    \includegraphics[scale=0.9]{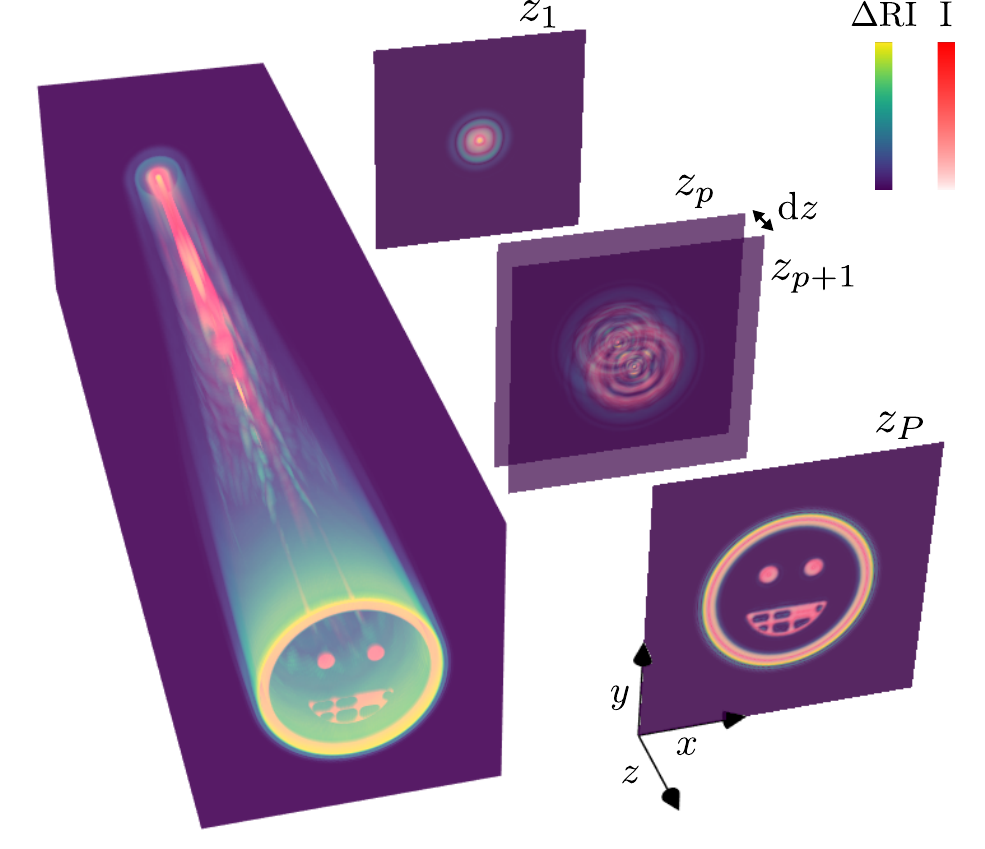}
    \caption{Illustration of a Gaussian to smiley gradient index mode
      conversion along with its discrete formulation, consisting of
      $P$ planes separated by a small distance $\mathrm{d}z$.}
    \label{fig:design}
\end{figure}

We consider a refractive index distribution $\Delta\textrm{RI}(x,y,z)$
of finite extent in a rectangular coordinate system ($O,x,y,z$), $z$
being the direction of propagation of the incoming laser modes
$\{u_n\}_{1\leq n \leq N}$. The wavelength of the input and output
modes in vacuum is denoted $\lambda_0$. The RI volume is discretized
into $P$ transverse planes located at axial positions
$\{z_p\}_{1\leq p \leq P}$ separated by a small distance
$\mathrm{d}z$, as illustrated in Figure~\ref{fig:design}. The
transverse resolution is denoted $(\mathrm{d}x,\mathrm{dy})$. The
refractive index $n_b$ of the medium between these different planes is
assumed to be the same as for the bulk material, while each plane
holds a phase mask proportional to the local $\Delta\textrm{RI}$
distribution and to $\mathrm{d}z$:
\begin{equation}
    \Delta\varphi_p(x,y) = \frac{2\pi}{\lambda_0}\Delta\textrm{RI}(x,y,z_p)\mathrm{d}z,
    \label{eq:phase_mask}
\end{equation}
for $p\in\{1,\dots,P\}$. Then, the numerical propagation of the input
modes $\{u_n\}$ through the $\Delta\textrm{RI}$ distribution is
achieved with the standard split-step beam propagation method (BPM),
alternating propagation in a uniform medium of refractive index $n_b$
and multiplication by a phase mask defined according to
Eq.~(\ref{eq:phase_mask}). A symmetric split-step scheme is enforced
simply by defining the input plane position $z_{in}$ at a distance
$\mathrm{d}z/2$ before $z_1$ and the output plane position $z_{out}$
at $\mathrm{d}z/2$ after $z_P$.

For propagating between planes we use the well-known angular spectrum
(AS) method:
\begin{equation}
    \begin{aligned}
    u_{\perp(z+\textrm{d}z)} & = \textsc{IFFT}\left(
    \textsc{FFT}{\left(u_{\perp z}\right)}\times\exp{\left(i 2\pi \mathrm{d}z \sqrt{\frac{n_b^2}{\lambda_0^2}-f_x^2-f_y^2}\right)}\right) \\
    \end{aligned}
\label{eq:AS}
\end{equation}
where \textsc{FFT} and \textsc{IFFT} refer to the two-dimensional fast
Fourier transform and its inverse. {A detailed pseudocode description
  of the forward model, propagating a N-vector of input modes $U_{in}$
  defined in the input plane $z_{in}$ to a N-vector of output modes
  $U_{out}$ in the destination plane $z_{out}$, is given in
  Algorithm~\ref{alg:forward_model} of Appendix~\ref{appendix:algorithms}.}

\subsection{Cost functions and errors for different classes of problems}

Depending on the beam shaping application, different cost functions
can be defined{, which quantify the quality of the light transform by
  a single scalar value $C$}. Here, we consider two types of cost
functions, either distances that we want to minimize to zero, or
functions of merit representing a physical quantity that we want to
maximize.

In order to keep the notations simple, we still name
$\{u_n\}_{1\leq n\leq N}$ the elements of the N-vector $U_{out}$
representing the input modes propagated to the plane $z_{out}$. We
note $\bar{U} = (\partial C/\partial u_n)_{1\leq n \leq N}$ the error
vector whose elements $\{\bar{u}_n\}$ represent the variation of the
cost function with respect to each propagated mode $u_n$. The
$\{u_n\}$ being complex-valued, it is important to define some
consistent algebraic rules for achieving such differentiation. We use
the complex representation defined in~\cite{Jurling:14} along with the
corresponding differentiation rules. It is worth noting that other
representations and derivation rules based on Wirtinger calculus can
be used~\cite{chakravarthula2019wirtinger}, but they should lead to
the same result in the end when error computations with respect to
real-valued parameters (e.g. phases) are performed. {Using such rules,
  we see that the errors $\{\bar{u}_n\}$ have the same dimension as
  the input and output modes $\{u_n\}$ and $\{v_n\}$, so we call them
  error modes in the following.}  Table~\ref{table:cost_functions}
lists a few interesting cost functions {and associated error modes}
for different beam shaping problems, which we describe in the
following.

\begin{table}[ht]
\centering
\begin{tabular}{|l|l|l|}
\hline
\textbf{beam shaping class} & \textbf{cost function C} & \textbf{error} $\mathbf{\bar{U}}$ \\
\hline
multimode matching & \(\min \sum\limits_{n=1}^N\iint\vert u_n - v_n \vert^2\) &
\(\bar{u}_n = 2(u_n-v_n)\) \\
\hline
multimode 1--1 power coupling &
\(\max \sum\limits_{n=1}^{N}{\left\vert\iint{v_n^*u_n}\right\vert^2}\) &
\(\bar{u}_n = 2\left(\iint{v_n^*u_n}\right)v_n\) \\
\hline
multimode N--N power coupling &
\(\max \sum\limits_{n=1}^{N}\sum\limits_{l=1}^{N}{
\left\vert\iint{v_l^*u_n}\right\vert^2}\) &
\(\bar{u}_n = 2\sum\limits_{l=1}^N{\left(\iint{v_l^*u_n}\right)}v_l\) \\
\hline
multimode intensity shaping &
\(\min \iint\left(I_S-I_T\right)^2\) & \(\bar{u}_n = 4(I_S-I_T)u_n\) \\
\hline
\end{tabular}
\caption{Examples of cost functions and associated errors for different classes of beam shaping problems.}
\label{table:cost_functions}
\end{table}

The first cost function we refer to as "multimode matching" is a
squared $l_2$ distance between the modes $\{u_n\}$ and $\{v_n\}$ with
a one-to-one correspondence. We call it that way because, in the case
of lossless designs, its application matches rigorously the wavefront
matching method introduced in a previous
article~\cite{sakamaki2007new} in the context of mode multiplexing
with planar lightwave circuits (PLC). Indeed, a lossless design
implies the conservation of $\|u_n\|_2^2$ during $u_n$ propagation,
which leads to a simpler expression for the error vector components:
\begin{equation}
    \bar{u}_n = -2v_n.
    \label{eq:wavefront_matching}
\end{equation}
We see that, up to a constant factor, the error modes correspond to
the target modes $\{v_n\}$.  {In this case, the application of the
  backpropagation of errors we develop in the next subsection leads
  exactly} to the wavefront matching method presented
in~\cite{sakamaki2007new}, even if the authors obtain this result with
an alternative reasoning.

If the {phase offset} between each mode pair $u_n$ and $v_n$ is not
relevant for the multiplexing problem, a different cost function can
be introduced that we name "multimode 1--1 power coupling". This time,
the cost function represents the sum of overlap integrals between
input and target modes and must therefore be maximized. We note that
the error vector components differ from the simplified expression of
multimode matching given in Eq.~(\ref{eq:wavefront_matching}) only by
a complex constant factor corresponding to the overlap integrals and
by a sign which just results from the difference between minimization
and maximization problems. This slight modification pre-aligns the
relative constant phases between each $u_n$ and $v_n$, which can
sometimes help to reduce the inverse design complexity.

We introduce a third cost function named "multimode N--N power
coupling", which aims at maximizing the total power coupling between
the input set of modes $\{u_n\}$ and the target modes $\{v_n\}$
\emph{without} requiring a one-to-one mapping. This metric can be
useful for applications such as incoherent combining of a set of modes
into a multimode fiber, or for space division multiplexing (SDM) in
telecommunications~\cite{puttnam2021space}, as we will illustrate in
the numerical results section. Compared to the multimode 1--1 power
coupling metric, this one gives much more flexibility to the
optimization algorithm since it allows {for mapping each input mode
  onto an arbitrary linear combination of target modes
  $\{v_n\}$. Usually, this additional degree of freedom results in
  smoother and more symmetrical designs}.

Finally, we introduce a cost function named "multimode intensity shaping", which allows to shape the total intensity distribution $I_S$ of mutually incoherent input modes $\{u_n\}$ into a desired target intensity shape $I_T$. For this particular beam shaping application, it is not necessary to define a set of target modes $\{v_n\}${, as only $I_T$ is relevant}. Moreover, the $\{u_n\}$ being mutually incoherent, their time-averaged total intensity profile is simply given by:
\begin{equation}
    I_S = \sum\limits_{n=1}^N{\vert u_n\vert^2}.
\end{equation}

In the following, we describe the backpropagation algorithm that
allows to compute the gradients $\partial C/\partial\Delta\textrm{RI}$
for any cost function $C$ in a unified fashion, starting from its
associated and {case specific} error vector $\bar{U}$.

\subsection{Backpropagation of errors}

Algorithm~\ref{alg:gradient_computation} of
Appendix~\ref{appendix:algorithms} describes the backpropagation of
the error modes vector $\bar{U}$ in order to compute the refractive
index gradients associated to each plane
$\nabla_\textrm{RI}[p] = \partial C/\partial\Delta\textrm{RI}[p]$ in a
reverse fashion. Here, the term "backpropagation" refers to the
traditional definition found in machine learning, where the chain rule
is applied in order to compute the partial derivatives of the cost
function with respect to some parameters of interest
backwards. Following the complex representation of errors defined
in~\cite{Jurling:14}, it appears that this procedure also corresponds
exactly to the physical backpropagation of the error modes through the
optical system.

\subsection{Optimization algorithm}

The optimization algorithm relies on gradient descent and can be summarized in the following steps:
\begin{enumerate}
    \item Initialize $\Delta\textrm{RI}$.
    \item Compute the propagated modes $U_{out}$ by calling the
      function \textsc{propagate\_forward}
      (Algorithm~\ref{alg:forward_model}) on the input modes $U_{in}$.
    \item Eval $C$ and $\bar{U}$, associated to a particular beam shaping problem, using the propagated modes $U_{out}$.
    \item Compute the refractive index gradients $\nabla_\textrm{RI}$
      by calling \textsc{backpropagate\_gradient}
      (Algorithm~\ref{alg:gradient_computation}) on the error modes
      $\bar{U}$.
    \item Update $\Delta\textrm{RI}$ with a gradient step along $\nabla_\textrm{RI}$.
    \item Enforce constraints on $\Delta\text{RI}$.
    \item If $C$ satisfies a stopping criterion then return $\Delta\textrm{RI}$, otherwise go to step 2.
\end{enumerate}

Concerning step 1, we often initialize $\Delta\textrm{RI}$ to zero
even if it is worth mentioning that some better initialization schemes
could be used~\cite{kunkel2020numerical}. A clever initialization can
help to achieve a faster convergence in case the propagated modes
through a constant refractive index medium have a very low spatial
overlap with the target modes.

Concerning step 5, one needs to introduce a constant step size, small
enough to avoid divergence, or use an adaptive learning rate
method. Of course the direction of the gradient step has to be chosen
consistently, depending on whether the cost function has to be
minimized or maximized.

Concerning step 6, it is worth mentioning that enforcing constraints
right after performing gradient steps finds a mathematical
justification in the proximal algorithms
framework\cite{parikh2014proximal}. A typical constraint that we wish
to apply is the restriction of $\Delta\textrm{RI}$ to a fixed
interval, in order to satisfy manufacturing constraints for instance.

\section{Numerical results}\label{results}

In this section, we present some numerical results that illustrate the
versatility of the optimization algorithm in addressing different beam
shaping problems involving fiber optics. We aim at designing fully
integrated devices of only millimeter length, which do not require any
collimation or focusing of the input and output beams. The refractive
index contrast required to realize some of the proposed devices may be
difficult to achieve with current glass or polymer gradient
manufacturing processes, but the main aim of this work is to propose a
proof of concept and give an idea of the objectives to be reached so
that these types of devices can be industrially manufactured.

\subsection{Hermite-Gaussian mode sorter}

The first system we design is a Hermite-Gaussian (HG) mode sorter,
inspired by recent work~\cite{fontaine2019laguerre}, where the authors
perform a conversion of 210 modes from a set of Gaussian input beams
arranged in a triangle to a set of co-propagating HG modes.  {Mode
  sorters are a proposed solution for increasing the telecommunication
  bandwidth by exploiting the concept of selective mode-division
  multiplexing, i.e., coupling many signals into a few-mode or
  multimode fiber~\cite{gross2015ultrafast} with control of the
  particular excitation of a given output mode}.

In~\cite{fontaine2019laguerre}, the HG conversion is performed by only
7 phase masks separated by free space propagation and is implemented
experimentally by a multi-pass cavity formed by a spatial light
modulator (SLM) and a flat mirror. The algorithm used in
article~\cite{fontaine2019laguerre} is equivalent to a coordinate
descent formulation of the wavefront matching method presented
in~\cite{sakamaki2007new}. It is important to underline that this
complex transformation is only made possible by a clever one-to-one
mapping between the cartesian coordinates of a Gaussian mode in the
triangle arrangement and the order $(m,n)$ of the corresponding HG
output mode. With so few phase patterns, any other arrangement would
lead to poor convergence of the mode transformation.

\begin{figure}[ht]
  \centering \includegraphics[scale=1]{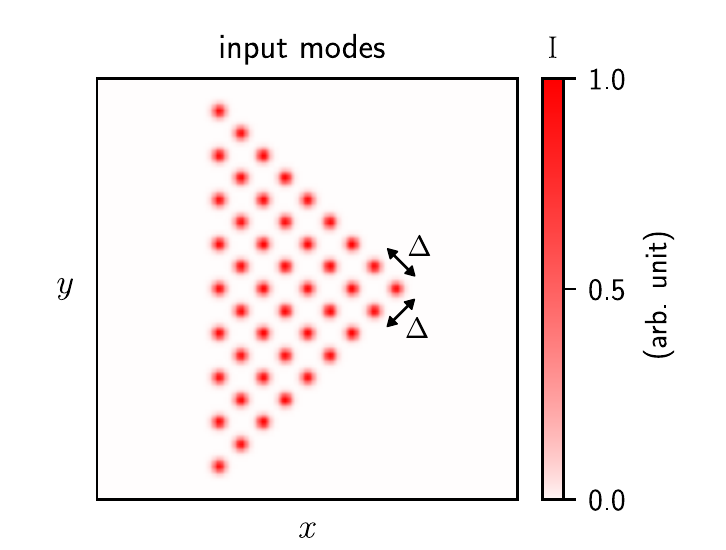}
  \caption{Triangle arrangement for the 45 Gaussian input modes.}
\label{fig:HG_sorter_input}
\end{figure}

In order to limit the computational resources to something reasonable,
we restrict the mode conversion to 45 modes, which still includes
interesting applications for
telecommunications~\cite{fontaine2018packaged}. We set the working
wavelength $\lambda_0 = 1.55\,\textrm{\textmu m}$ and use Gaussian
modes of waist $\omega_{in} = 5.2\,\textrm{\textmu m}$ as inputs,
typical of single mode fibers
(SMF-28). Figure~\ref{fig:HG_sorter_input} shows the triangle
arrangement of the 45 input modes, with a separation parameter
$\Delta = 4\,\omega_{in}$. For the outputs, we use the HG
representation of standard graded-index multimode fibers' eigenmodes,
with a maximum refractive index contrast
$dn_\textrm{max} = 15\times 10^{-3}$ between the core and the cladding
as in~\cite{fontaine2018packaged}. With a mode
solver~\cite{fallahkhair2008vector}, we compute the eigenmodes for
such a fiber profile and estimate the value
$\omega_{out} = 7.7\,\textrm{\textmu m}$ for the HG output modes'
waist. {To each of the 45 input modes, we assign one of the output
  fiber HG eigenmodes according to the mapping described
  in~~\cite{fontaine2019laguerre}. In short, if the Cartesian position
  of an input mode in the triangle arrangement (in a $45^\circ$
  rotated frame compared to Figure~\ref{fig:HG_sorter_input}) is
  expressed as $(m\Delta, n\Delta)$, then the associated output mode
  is $\textrm{HG}_{m,n}$.}

In simulation, the device is represented by a
$512\times 512\times 250$ array, the first two dimensions
corresponding to the transverse planes and the third one to the plane
index. The bulk refractive index used for the propagation between
planes is set to $n_b = 1.444$ corresponding to fused silica. The
transverse resolution is set to
$\mathrm{d}x = \mathrm{d}y = 0.65\,\textrm{\textmu m}$ and the
separation between planes is $\mathrm{d}z = 10\,\textrm{\textmu m}$,
leading to a total device length of 2.5 mm. At first glance, the axial
resolution may appear a bit low for BPM accuracy, but we verify
afterwards that it is feasible, by linearly interpolating the final
simulated RI profile with a better resolution
$\mathrm{d}z = 0.5\,\textrm{\textmu m}$ and by propagating the input
modes through it. Limiting the number of planes in the simulation is
crucial because we need to store all the input modes at each plane
during the forward pass (Algorithm~\ref{alg:forward_model}) in order
to be able to compute a gradient afterwards
(Algorithm~\ref{alg:gradient_computation}). Even with single-precision
complex numbers, this already requires 23.6 GB of memory allocation.

\begin{figure}[ht]
  \centering \includegraphics[scale=1]{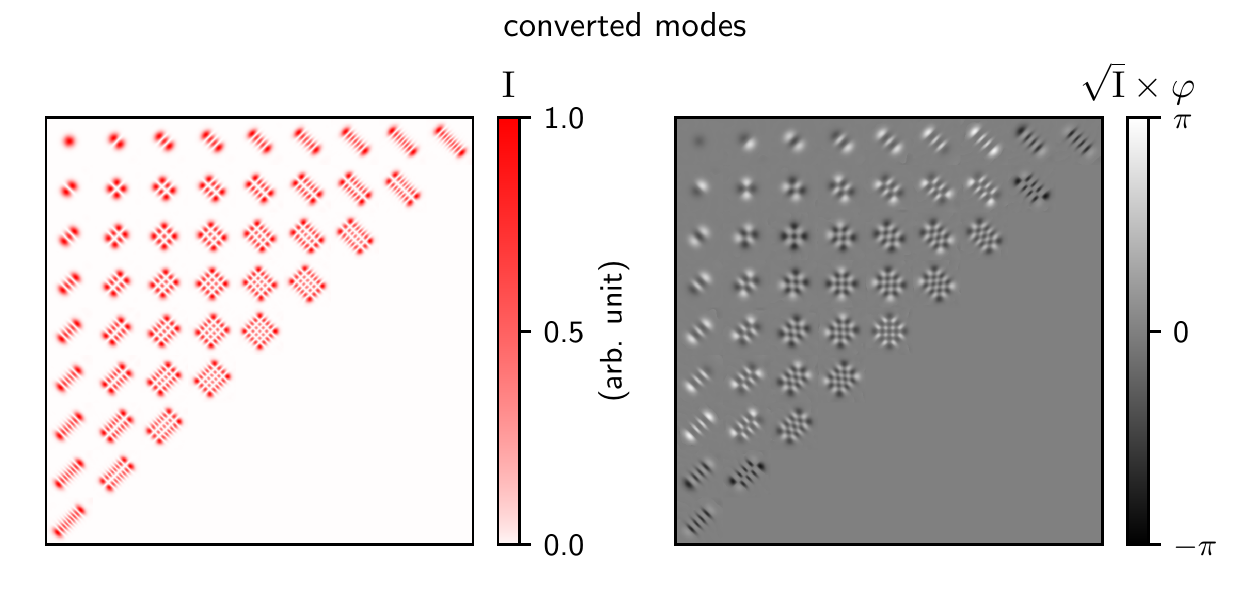}
    \caption{Intensity and phase patterns of the 45 created HG modes. The average conversion efficiency is 98.6\%.}
    \label{fig:HG_sorter_output}
\end{figure}

In order to achieve this 45-mode conversion, we run 1500 iterations of
the optimization algorithm using the multimode 1-1 power coupling cost
function defined in table~\ref{table:cost_functions}. After each
iteration, we force the values of $\Delta\textrm{RI}$ to stay within a
range of $12\times 10^{-3}$, which is slightly smaller than the RI
contrast $dn_\textrm{max}$ required to guide the output HG modes. We
observed that enforcing stricter constraints on $\Delta\textrm{RI}$
causes losses in the mode conversion due to the device's inability to
strongly guide the small modes involved. In the end of the iterations,
the cost function reaches the value $C \simeq 44.37$, corresponding to
an average conversion efficiency of 98.6\% per mode. As
in~\cite{fontaine2019laguerre}, we introduce the $N\times N$ transfer
matrix $T$ of the device, whose elements $t_{ij} = \int{v_i^*u_j}$ are
the overlaps between the output and the propagated input modes. For
this particular device, we find that the auto-correlation matrix
$X = \vert T^\dagger T\vert$ has a largest off-diagonal term
$X_\textrm{max} = -46.16$~dB, corresponding to an extremely small
crosstalk between modes. The intensity and phase of each converted
mode are shown in Figure~\ref{fig:HG_sorter_output}, where we
recognize almost perfectly shaped HG modes.

\begin{figure}[ht]
    \centering
    \includegraphics[scale=1]{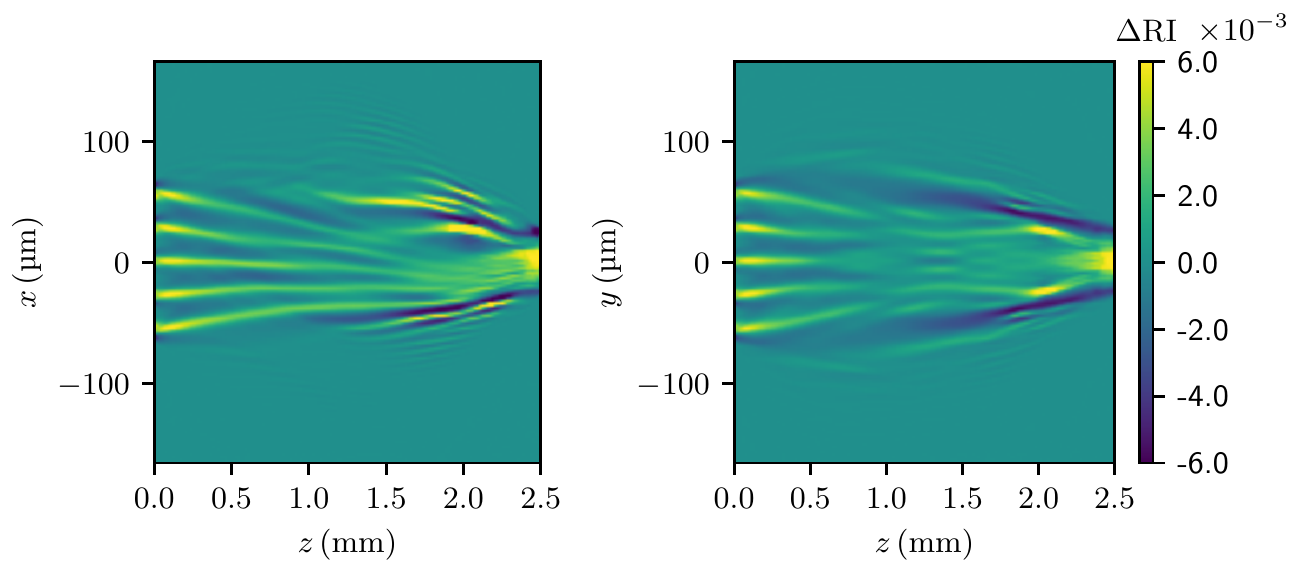}
    \caption{Evolution of the simulated transverse RI profile sections
      through the center of the mode sorter device along the
      propagation direction, for the multimode HG conversion.}
    \label{fig:HG_sorter_RI}
\end{figure}

Figure~\ref{fig:HG_sorter_RI} illustrates horizontal and vertical
sections {through the center of the mode converter, indicating a very
  smooth evolution of the refractive index along the propagation
  direction.} The input facet $z=z_{in}$ defines multiple waveguides
for each of the 45 Gaussian input modes while the end facet
$z=z_{out}$ defines a single larger waveguide holding the
co-propagating HG modes. In between, the waveguides are merged by the
algorithm in order to achieve the desired mode to mode mapping.

\subsection{Photonic lanterns}

Photonic lanterns are optical components allowing for a low-loss
conversion between a set of fundamental modes belonging to independent
single-mode waveguides and a set of high-order modes belonging to a
common multimode waveguide~\cite{birks2015photonic}.

{The functional difference between a photonic lantern and a mode
  sorter as discussed previously is that the lantern is an
  non-selective device that does not aim for a pre-defined
  mode-to-mode mapping. The only goal is to couple as much power into
  the set of modes supported by the output fiber, while preserving
  mode orthogonality.}  Such objects are usually fabricated by
tapering single-mode waveguides together or by femtosecond laser
writing and they find applications in
astrophotonics~\cite{norris2019astrophotonics} and
SDM~\cite{leon2014mode} for telecommunications.

The empirical adiabatic merging of different fiber cores gives rise to
super-modes by evanescent coupling, but it is challenging to ensure
that this process is lossless and that the final modal content exactly
matches a unitary superposition of modes belonging to a given few-mode
or multimode fiber. Assuming that we can control the RI distribution
as we wish, we will show that the multimode N-N power coupling metric
proposed in Table~\ref{table:cost_functions} is perfectly adapted to
overcome this difficulty. If the target waveguide was a graded-index
multimode fiber {whose eigenmodes are Hermite-Gaussian}, then the HG
mode sorter presented previously would already be an appropriate
solution. However, it is not necessarily possible to find a Cartesian
representation for the eigenmodes of an arbitrary fiber profile, {one
  example are the} linearly polarized (LP) modes of a step-index
fiber.

\begin{figure}[ht]
  \centering
  \includegraphics{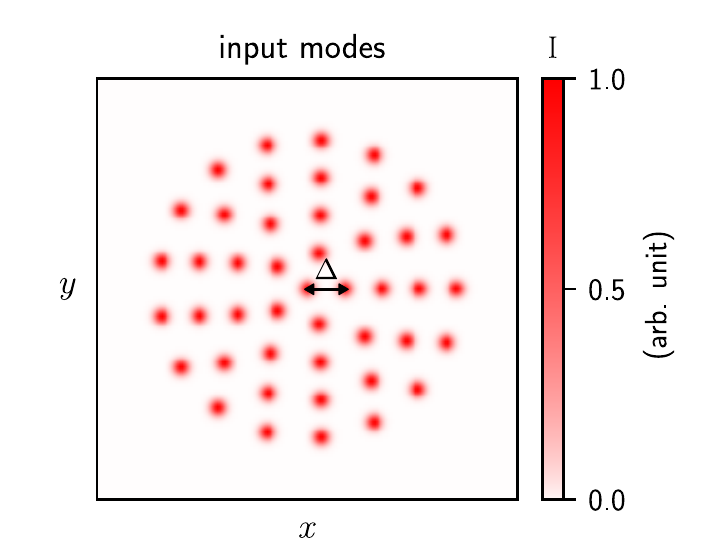}
  \caption{Concentric ring arrangement for the 45 input modes of the
    photonic lantern.}
  \label{fig:PL_input}
\end{figure}

Here, we still define the working wavelength
$\lambda_0 = 1.55\,\textrm{\textmu m}$ and the bulk refractive index
$n_b = 1.444$. The device is represented by a
$350\times 350 \times 250$ array, with the same transverse and axial
resolution as before, leading again to a total device length of 2.5
mm. Concerning the input modes, we still use the fundamental Gaussians
of single-mode fibers with $\omega_{in} = \textrm{5.2 \textmu m}$, but
this time with a concentric ring arrangement which has been shown to
lead to the best coupling efficiency for a photonic lantern design
with tapered
waveguides~\cite{fontaine2012geometric}. Figure~\ref{fig:PL_input}
represents the concentric ring arrangement for 45 input modes, with a
separation parameter $\Delta = 4\,\omega_{in}$ as before.

\begin{figure}[ht]
  \centering
  \includegraphics{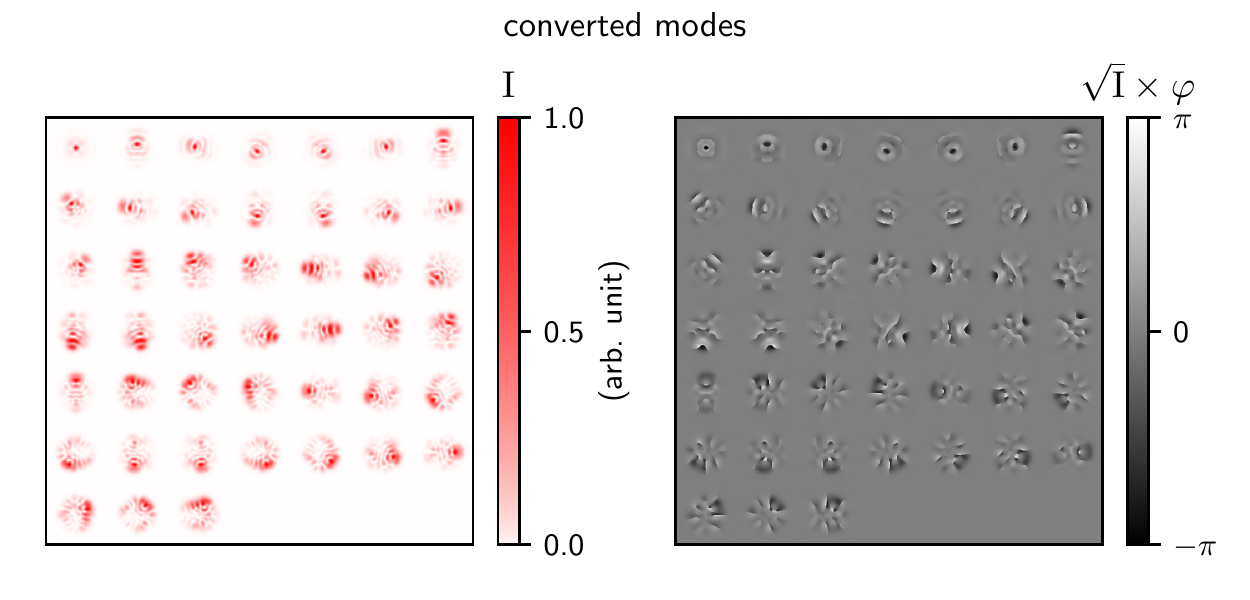}
  \caption{Intensity and phase profiles of the converted modes.}
  \label{fig:PL_converted}
\end{figure}

Concerning the output modes, we use the first 45 Laguerre-Gaussian
(LG) modes with $\omega_{out} = \textrm{7.7 \textmu m}$, using their
real-valued representation with circular and radial nodal lines. For
the multimode N-N power coupling optimization, the exact base
representation is not relevant since it allows for a linear
rearrangement of the modes, but we find it nicer for visualizing the
intensity profiles later on.

After only 300 iterations of the optimization algorithm, we reach a
total power coupling efficiency of 99.2\%. As
in~\cite{fontaine2019laguerre}, we define insertion losses (IL) and
mode dependent losses (MDL) of the device using the singular values
$\{\sigma_i\}_{1\leq i \leq N}$ resulting from the singular value
decomposition (SVD) of the transfer matrix $T = U\Sigma V^\dagger$:
\begin{align}
  \textrm{IL} = & \frac{1}{N}\sum\limits_{i=1}^N\vert \sigma_i\vert^2 \\
  \textrm{MDL} = & \frac{\max{\{\vert \sigma_i\vert^2\}}}{\min{\{\vert \sigma_i \vert^2\}}}.
\end{align}

\begin{figure}[ht]
  \centering
  \includegraphics{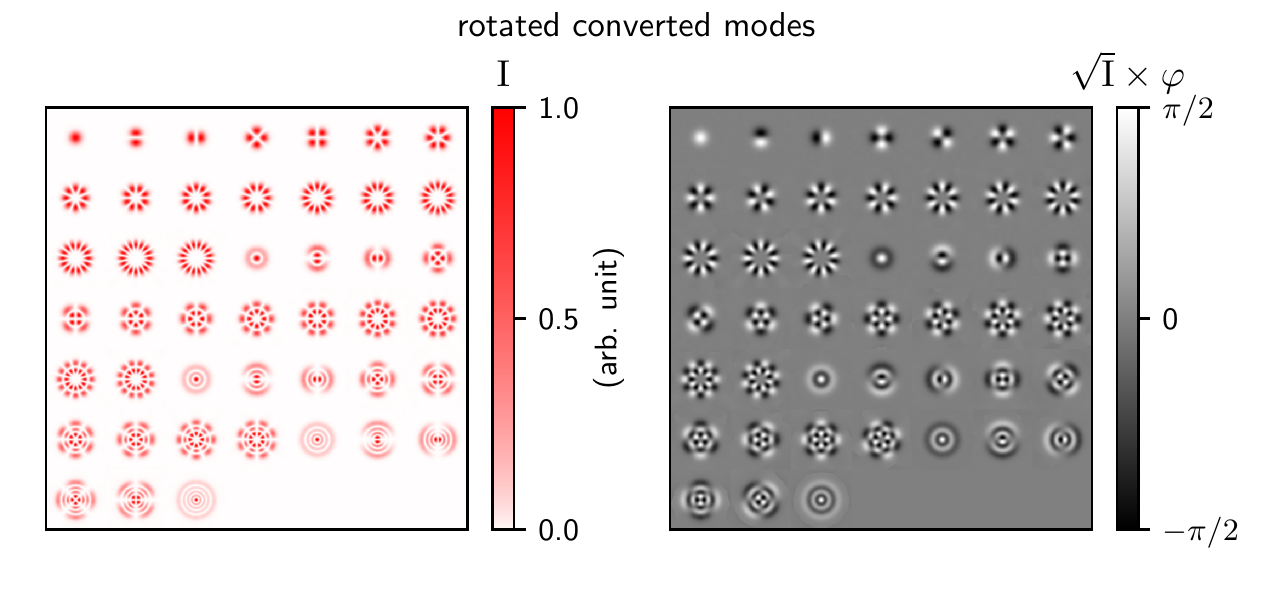}
  \caption{Intensity and phase profiles of the converted modes rotated by the unitary matrix $UV^\dagger$ resulting from the SVD.}
  \label{fig:PL_rotated}
\end{figure}

\begin{figure}[ht]
  \centering
  \includegraphics{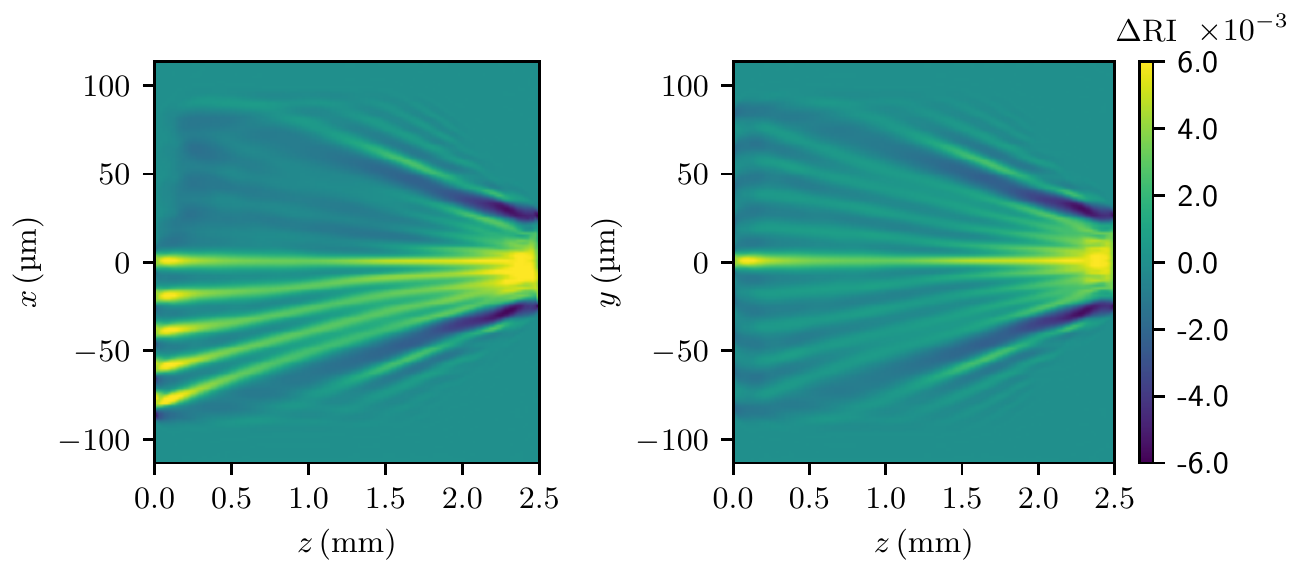}
  \caption{{Transverse RI profiles through the center of the photonic lantern design. The total power coupling efficiency is 99.2\%.}}
  \label{fig:PL_RI}
\end{figure}

For this particular device, we obtain $\textrm{IL} = 0.035$~dB,
$\textrm{MDL} = 0.097$~dB and a maximum crosstalk
$X_\textrm{max} = -46.5$~dB, corresponding to an auto-correlation
matrix $X = \vert T^\dagger T\vert$ almost equal to the identity
matrix. Figure~\ref{fig:PL_converted} shows that the converted modes
look very different from the LG modes defining the output basis, but
Figure~\ref{fig:PL_rotated} shows that, since
$\Sigma \simeq \mathbb{I}$, the original LG modes can be recovered by
rotating the converted modes with the unitary matrix $UV^\dagger$
resulting from the SVD decomposition of the transfer matrix $T$.

Finally, Figure~\ref{fig:PL_RI} shows that the computed design is very
smooth and symmetrical, resembling a tapering of several single-mode
waveguides, thus truly deserving the name photonic lantern. Our
results prove that our method is able to engineer integrated photonic
lantern devices in a very precise fashion, with perfectly controlled
input and output modes. To our knowledge, this inverse design approach
is new in this context.

\subsection{Multimode intensity shaping}

{Finally, we would like to treat the application of shaping the
  intensity profile of a light source emitting many mutually
  incoherent modes.  This class of beam shaping has importance in
  various fields, such as digital holography or laser material
  processing, where the native intensity profile of the light source
  is often not ideal for the task at hand.  } The associated cost
function is called ``multimode intensity shaping'' in
Table~\ref{table:cost_functions}. In a recent article, we have used
the same optimization metric to find two cascaded phase patterns for
shaping the intensity profile of a light emitting diode
(LED)~\cite{barre2021holographic}. Here, we aim at realizing the same
kind of intensity shaping with a volumetric gradient index design,
starting from the eigenmodes of a step-index fiber with a diameter of
$50\,\textrm{\textmu m}$ and $\textrm{NA} = 0.1$.

In simulation, we use $\lambda_0 = 0.640\,\textrm{\textmu m}$ as the
working wavelength and $n_b = 1.457$ as the refractive index of fused
silica at this wavelength. With such parameters, a mode solver finds
158 eigenmodes for this particular step-index fiber.  {We further
  assume that each input mode contributes with the same optical
  power.} The device is represented by a $256\times 256\times 50$
array with a transverse resolution
$\mathrm{d}x = \mathrm{d}y = 0.5\,\textrm{\textmu m}$ and a separation
between planes $\mathrm{d}z = 10\,\textrm{\textmu m}$. This
corresponds to a total device length of only 0.5~mm.  {As previously,
  we validate our sampling choice by linearly interpolating the final
  RI profile along the z-direction at a finer sampling interval of
  $\mathrm{d}z = 0.5\,\textrm{\textmu m}$, followed by a simulated
  readout of the gradient index design.}

As desired output, we define a square target intensity profile $I_T$
with a side length of $60\,\textrm{\textmu m}$. The square is modeled
as the Cartesian product of two 1D supergaussian profiles of order 20
to provide smooth edges.

\begin{figure}[ht]
  \centering
  \includegraphics{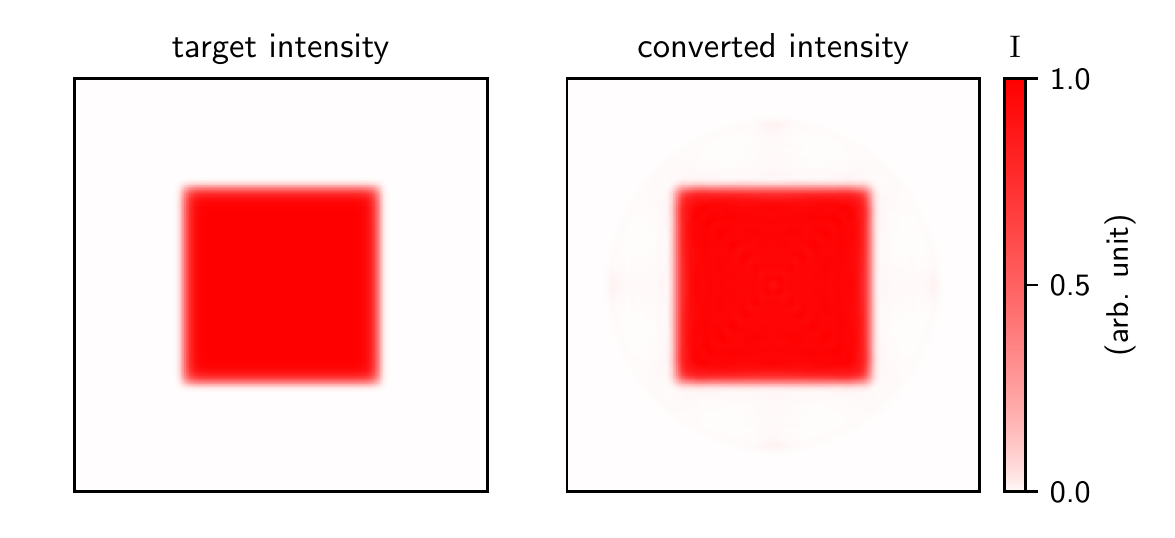}
  \caption{Theoretical target intensity profile and converted multimode intensity profile obtained by simulation.}
  \label{fig:square_shaping}
\end{figure}

\begin{figure}[ht]
  \centering
  \includegraphics{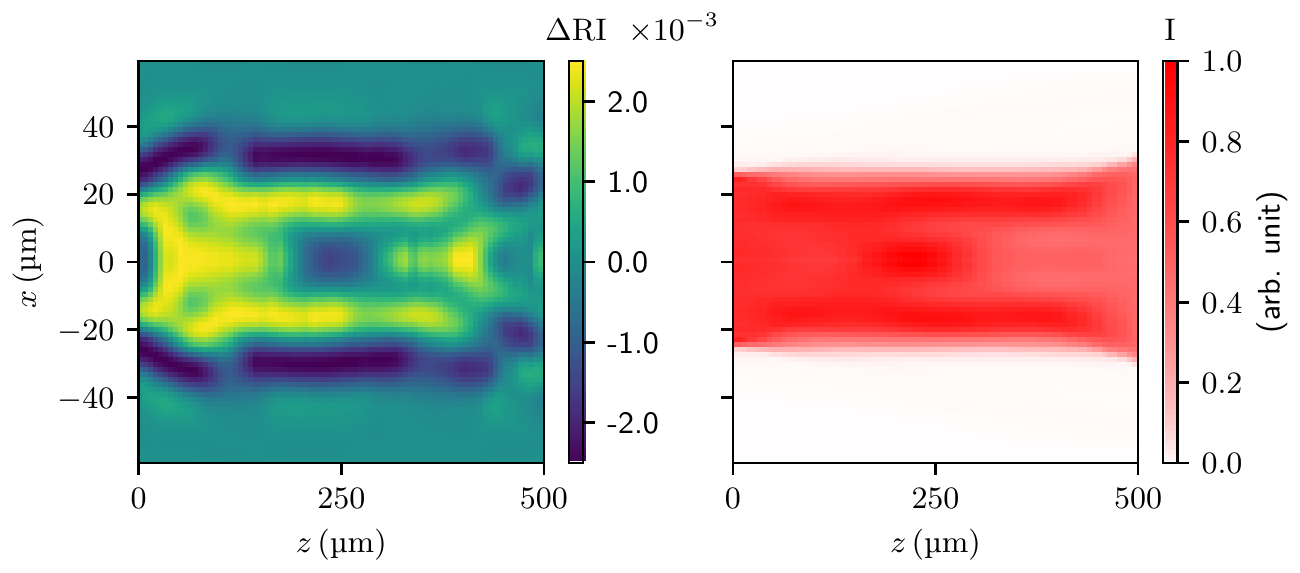}
  \caption{{RI profile and total intensity profile in the central x-z-plane of the beam shaper.}}
  \label{fig:square_transform}
\end{figure}

For this simulation, we restrict the $\Delta\textrm{RI}$ range to
$5\times 10^{-3}$ and we also enforce a smooth design by Fourier
filtering the 3D RI profile with a sharp low-pass filter at each
iteration. This Fourier filtering corresponds to a projection of the
RI distribution on a set of band-limited functions, which makes it a
theoretically valid proximal operator~\cite{parikh2014proximal}. Here,
we enforce the {smallest details} in the RI profile to be around
$4\,\textrm{\textmu m}$. After 1500 iterations, even under these
constraints, we reach an almost perfect intensity conversion as
illustrated in Figure~\ref{fig:square_shaping}.

Figure~\ref{fig:square_transform} shows the RI profile and the total
intensity evolution of the beam along the propagation direction. We
observe that the obtained RI profile is very smooth, and its limited
$\Delta\textrm{RI}$ range makes it already manufacturable with
commercial 3D printers with polymers~\cite{Porte:21}, even if it may
require some slice stitching in order to reach the full 0.5~mm
length. We also see that the beam is guided in the computed RI
structure while its intensity is continuously shaped into the desired
profile. The fact that the total length is 5 times shorter for this
multimode shaping device than for the previous mode sorter or photonic
lantern shows that multimode intensity shaping requires in general
much less degrees of freedom than individual or collective mode
matching. The same conclusion holds for discrete designs such as
presented in Refs.~\cite{barre2021holographic}
and~\cite{fontaine2019laguerre}.

\section{Conclusion}

We have demonstrated a powerful and versatile computational method to
design millimeter-range integrated devices performing several
important multimode light transformations.  Different cost functions
were introduced for solving particular mode multiplexing or intensity
shaping problems, but the optimization approach is stated in such a
general way that any other user-defined cost function would fall
within the scope presented here.  The presented designs can already be
manufactured with the available two-photon polymerization technologies
and we can envision that their realization in a glass material will
also be accessible in the coming years. Our work could be extended
further to include broad-spectrum or multicolor sources, or to account
for light polarization, which would cover many more applications in
optics with fully integrated designs.

\clearpage

\appendix

\section{Algorithms}\label{appendix:algorithms}

Both forward and backward propagation rely on the angular spectrum
method which is implemented by the
$\textsc{as\_prop}(u, \lambda_0, n_b, \mathrm{d}z)$ function, $u$
being the transverse mode to propagate, $\lambda_0$ the wavelength in
vacuum, $n_b$ the bulk refractive index and $\mathrm{d}z$ the
propagation distance.

\subsection{Forward model}

\begin{algorithm}[ht]
\caption{Forward model}\label{alg:forward_model}
\begin{algorithmic}
\State $\mathbf{Constants:}$ $\lambda_0$, $n_b$, $\mathrm{d}z$
\State $\mathbf{Inputs:}$ $U_{in}$ a $N$-vector of input modes at plane $z_{in}$\\
\phantom{$\mathbf{Inputs:}$} $\Delta\mathrm{RI}$ a $P$-vector of 2D refractive index distributions
\State $\mathbf{Outputs:}$ $U_{out}$ a $N$-vector of modes propagated to plane $z_{out}$\\
\phantom{$\mathbf{Outputs:}$} $S$ a $P$-vector of $N$-vectors of modes stored at each plane
\Procedure{propagate\_forward}{$U_{in}$, $\Delta\mathrm{RI}$}
\State $S \gets \Call{Vector}{\mathrm{typeof}(U_{in}), P}$
\Comment allocating a P-vector of N-vectors of modes
\State $U_{out} \gets \Call{copy}{U_{in}}$ \Comment{copy allocation}
\State $U_{out} \gets \Call{as\_prop.}{U_{out}, \lambda_0, n_b, \mathrm{d}z/2}$
\Comment{propagating modes to plane $z_1$}
\State $U_{out} \gets U_{out}\cdot\exp{(i \frac{2\pi}{\lambda_0} \Delta\mathrm{RI}[1])}$
\Comment{crossing the $1$\textsuperscript{st} refractive index mask forward}
\State $S[1] \gets U_{out}$ \Comment{storing modes at plane $z_1$}
\For{$p=2\,\mathbf{to}\,P$}
  \State $U_{out} \gets \Call{as\_prop.}{U_{out}, \lambda_0, n_b, \mathrm{d}z}$
  \Comment{propagating modes to the next plane}
  \State $U_{out} \gets U_{out}\cdot\exp{(i \frac{2\pi}{\lambda_0} \Delta\mathrm{RI}[p])}$
  \Comment{crossing $p$\textsuperscript{th} refractive index mask forward}
  \State $S[p] \gets U_{out}$ \Comment{storing modes at plane $z_p$}
\EndFor
\State $U_{out} \gets \Call{as\_prop.}{U_{out}, \lambda_0, n_b, \mathrm{d}z/2}$
\Comment{propagating modes to plane $z_{out}$}
\State \Return $(U_{out},S)$
\EndProcedure
\end{algorithmic}
\end{algorithm}

Algorithm~\ref{alg:forward_model} describes the forward model in a
more formal way. The set of input modes is represented by an N-vector
$U_{in}$, each of its elements being a two-dimensional array storing a
particular mode distribution $\{u_n\}_{1\leq n \leq N}$ in plane
$z_{in}$. The 3D refractive index distribution is represented by a
P-vector $\Delta\textrm{RI}$ whose elements are two dimensional arrays
representing the refractive index profiles at each plane $z_p$. The
function \textsc{propagate\_forward} propagates the vector of modes
$U_{in}$ through the 3D refractive index distribution
$\Delta\textrm{RI}$, plane by plane, and returns a new vector
$U_{out}$ of propagated modes to the final plane $z_{out}$. It also
returns $S$, a stored copy of each input mode at each plane, which
will be crucial for computing the refractive index gradients later
on. In order to avoid verbose loops over the mode index, we introduced
a broadcasting notation by appending a \emph{dot} (.) after
\textsc{as\_prop} function name, meaning that we apply this function
to each element of the vector passed as first argument, and that the
return type is also a vector of same length. Similarly, the pointwise
multiplication of each vector component by a common phase mask has
been represented by a \emph{dot} $(\cdot)$ operator. Finally, it is
worth mentioning that these vector representations for modes and RI
distributions are useful for a concise and functional pseudo-code
description, but that in practice they should rather be represented by
contiguous multidimensional arrays for better efficiency.

\subsection{Error backpropagation}

\begin{algorithm}[ht]
\caption{RI gradient computation}\label{alg:gradient_computation}
\begin{algorithmic}
\State $\mathbf{Constants:}$ $\lambda_0$, $n_b$, $\mathrm{d}z$
\State $\mathbf{Inputs:}$ $\bar{U}$ a N-vector of error modes associated to a cost function at plane $z_{out}$ \\
\phantom{$\mathbf{Inputs:}$} $S$ a $P$-vector of $N$-vectors of modes stored at each plane\\
\phantom{$\mathbf{Inputs:}$} $\Delta\mathrm{RI}$ a $P$-vector of 2D refractive index distributions
\State $\mathbf{Outputs:}$ $\nabla_\mathrm{RI}$ a P-vector of RI gradients
\Procedure{backpropagate\_gradient}{$\bar{U}$, $S$, $\Delta\mathrm{RI}$}
\State $\nabla_\mathrm{RI} \gets \Call{similar}{\Delta\mathrm{RI}}$
\Comment allocating an uninitialized RI gradient vector
\State $\bar{U} \gets \Call{as\_prop.*}{\bar{U}, \lambda_0, n_b, \mathrm{d}z/2}$
\Comment{backpropagating errors to plane $z_P$}
\For{$p=P\,\mathbf{to}\,2$}
\State $\nabla_\mathrm{RI}[p] \gets \frac{2\pi}{\lambda_0}\Im{\left(\sum\limits_{k=1}^{N}(S^*[p]\cdot \bar{U})[k]\right)}$ \Comment{Computing the $p$\textsuperscript{th} RI gradient}
\State $\bar{U} \gets \bar{U}\cdot\exp{(-i \frac{2\pi}{\lambda_0} \Delta\mathrm{RI}[p])}$
\Comment{crossing the $p$\textsuperscript{th} refractive index mask backward}
\State $\bar{U} \gets \Call{as\_prop.*}{\bar{U}, \lambda_0, n_b, \mathrm{d}z}$
\Comment{backpropagating errors to the previous plane}
\EndFor
\State $\nabla_\mathrm{RI}[1] \gets \frac{2\pi}{\lambda_0}\Im{\left(\sum\limits_{k=1}^{N}(S^*[1]\cdot \bar{U})[k]\right)}$ \Comment{Computing the $1$\textsuperscript{st} RI gradient}
\State \Return $\nabla_\mathrm{RI}$
\EndProcedure
\end{algorithmic}
\end{algorithm}

The function \textsc{backpropagate\_gradient} can be seen as a
mirrored version of the function \textsc{propagate\_for\-ward} defined
in algorithm~\ref{alg:forward_model}, where each operator has been
replaced by its adjoint (hermitian conjugate). The \emph{star} ($^*$)
appended after the \textsc{as\_prop} function name denotes this
adjoint operation. Additionally, we see that the gradient computations
$\nabla_\textrm{RI}[p]$ are performed using the information $S[p]$
stored in each plane during the forward pass. These gradient
expressions can be derived from the set of rules presented
in~\cite{Jurling:14}.

In case of a tight grid extension, which is always desirable in order
to save memory consumption, some Fourier artifacts can easily appear
due to a portion of unguided light reaching the edges. This problem
can be overcome simply by adding some absorbing boundary conditions on
a thin region before the grid edges. In the forward model, this can
for instance be implemented by applying an absorption mask on the
propagating modes in conjunction with the phase mask
multiplication. It is worth mentioning that in the backward pass, it
must then also be treated as an absorption mask for the
backpropagating modes, as the adjoint operator needs to be considered
(and not the inverse).

\vfill

\paragraph{Funding.}
The work is funded by the FWF (I3984-N36).

\clearpage

\let\OLDthebibliography\thebibliography
\renewcommand\thebibliography[1]{
  \OLDthebibliography{#1}
  \setlength{\parskip}{1ex}
  \setlength{\itemsep}{0pt plus 1ex}
}

\bibliographystyle{unsrt}
\bibliography{gradient_design}

\end{document}